\documentclass[prb,twocolumn,floatfix,superscriptaddress]{revtex4}
\usepackage{amsmath}
\usepackage[dvips]{graphicx}
\usepackage{float}

\begin{document}

\title{The Shapes of Cooperatively Rearranging Regions in Glass Forming
Liquids}
\date{\today}
\author{Jacob D. Stevenson}
\affiliation{Department of Physics and Department of Chemistry and Biochemistry,
University of California, San Diego, La Jolla, CA 92093}
\author{J\"{o}rg Schmalian}
\affiliation{Department of Physics and Astronomy and Ames Laboratory, Iowa State
University, Ames, IA 50011}
\author{Peter G. Wolynes}
\affiliation{Department of Physics and Department of Chemistry and Biochemistry,
University of California, San Diego, La Jolla, CA 92093}

\begin{abstract}
The shapes of cooperatively rearranging regions in glassy liquids change from being compact at low temperatures to fractal or ``stringy'' as the dynamical crossover temperature from activated to collisional transport is approached from below. We present a quantitative microscopic treatment of this change of morphology within the framework of the random first order transition theory of glasses. We predict a correlation of the ratio of the dynamical crossover temperature to the laboratory glass transition temperature, and the heat capacity discontinuity at the glass transition, $\Delta C_p$. The predicted correlation agrees with experimental results for the 21 materials compiled by Novikov and Sokolov.
\end{abstract}

\maketitle

Our increased ability to visualize and experimentally probe supercooled
liquids on the nanometer length scale has explicitly revealed the presence
of cooperatively rearranging regions\cite%
{russell.2000,sillescu.1999,richert.2002,deschenes.2001,lacevic.2004,gebremichael.2004,reinsberg.2002,ediger.2000,weeks.2000,kob.1997}
(CRR's). The cooperative rearrangement of groups of many molecules has long
been thought to underlie the dramatic slowing of liquid dynamics upon
cooling and could also explain the non-exponential time dependence of
relaxation in glassy liquids. Activated transitions of regions of growing
size were postulated in the venerable Adam-Gibbs argument for the glass
transition\cite{adam.1965}. To move, a region, in the AG view, must have a
minimum of two distinct conformational states. Natural as this suggestion
is, the sizes predicted from literally applying this notion are far too
small to explain laboratory observations. The minimal AG cluster would have
only two particles since the measured entropy per particle is of order $1k_B$
at the glass transition.

A distinct approach, the random first order transition (RFOT) theory of
glasses, is based on a secure statistical mechanical formulation at the mean
field level\cite%
{singh.1985,kirkpatrick.1987a,kirkpatrick.1987b,kirkpatrick.1987c,mezard.1999,franz.2005a,franz.2005b,bouchaud.2004}
but also goes beyond mean field theory to explain the non-exponential,
non-Arrhenius dynamics of supercooled liquids through the existence of
compact, dynamically reconfiguring regions (``entropic droplets'')\cite%
{kirkpatrick.1989,xia.2000,xia.2001} whose predicted size is, in contrast to
the Adam-Gibbs bound, very much consistent with what has been measured
(125-200 molecules), using both scanning microscopy\cite%
{russell.2000,deschenes.2001} and NMR techniques\cite%
{sillescu.1999,richert.2002}, at temperatures near to $T_g$.

\begin{figure}[tbp]
\includegraphics[width=0.48\textwidth]{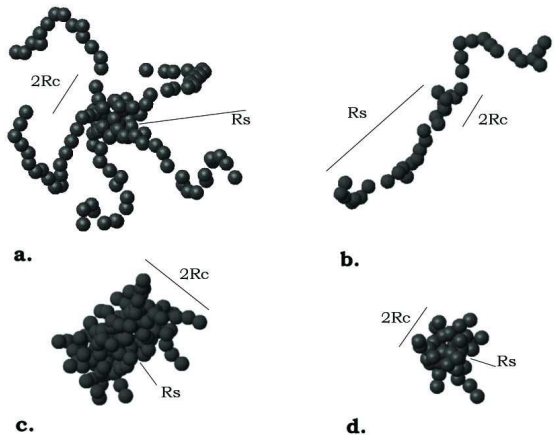}
\caption{The schematic appearance of the reconfiguring regions predicted by
RFOT theory according to the free energy profiles of the fuzzy sphere model
(see text) at: \textbf{a}. the final state near $T_c^{string}$, \textbf{b}.
the transition state near $T_c^{string}$, \textbf{c}. the final state near $%
T_g$, \textbf{d}. the transition state near $T_g$. The labeled lines
indicate the radii of the core, $R_c$, and of the fuzzy halo, $R_s$.}
\label{fig.1}
\end{figure}

Computer simulations\cite{gebremichael.2004,donati.1998,donati.1999} and
light microscopy studies of colloidal glasses\cite{weeks.2000}, however,
have revealed dynamically reconfiguring regions that are not compact and
contain fewer particles. Some investigators describe these regions as
``fractal\cite{reinsberg.2002}'' while others use the term ``strings\cite%
{gebremichael.2004}'' to characterize them. It has been suggested that such
``stringy'' excitations should be taken as the fundamental objects in the
theory of glass transitions. As we will show, the fractal nature of the
dynamically reconfiguring regions in the relatively high temperature regime
probed in current computer simulations follows naturally from RFOT theory.
To be precise, RFOT theory predicts the shape of the reconfiguring regions
changes from compact to fractal as the system is heated from low
temperatures, characteristic of the laboratory glass transitions, to the
higher dynamical crossover temperature, $T_A$, above which motions are no
longer activated. This transformation is shown in figure \ref{fig.1}. Owing
to the rather short time scale accessible to computer studies, simulations
have been inevitably carried out near this dynamical crossover. Likewise,
colloidal glasses in the laboratory are studied near to the dynamic
crossover because the large size of colloidal particles, in molecular terms,
means that the elementary constituents of such suspensions intrinsically
move slowly. The dimensionless time scales probed in experiments on colloids
are rather similar to those in computer studies.

According to RFOT theory, the dynamical crossover from activated motion has
the character of a spinodal\cite{kirkpatrick.1987b,biroli.2004}. Since an
analogous change of morphology of nucleation clusters is predicted to occur
in ordinary first order transitions\cite{unger.1984}, others have already
suggested that the dynamically heterogeneities near $T_A$ should be fractal
or percolation like\cite{johnson.1998}. Here we show how the RFOT theory
predicts the temperature range where the metamorphosis from compact to
fractal happens. We will show that according to RFOT theory, the gap between
the dynamical crossover temperature and the glass transition temperature for
molecular liquids should correlate inversely with the configurational heat
capacity. The existing experimental data confirm this prediction.

The mean field theory of random first order transitions starts by
constructing aperiodic minima of a free energy functional\cite{singh.1985}.
These aperiodic structures reflect density patterns of molecular units
vibrating in the vicinity of their current locations\cite%
{stoessel.1984,singh.1985,dasgupta.1999,fuchizaki.2002}. These aperiodic
free energy minima resemble the so-called ``inherent structures'' that are
minima of the potential energy, i.e. the latter are local minima of the $T=0$
free energy\cite{stillinger.1983}. At finite temperature, these aperiodic
structures represent a compromise between the cost of localizing a molecule $%
S_{loc}$ and the free energy gain realized by particles being able to get
out of each others way once localized. The latter free energy contribution
is represented by an interaction term in the usual free energy density
functional. Any resulting localized solution is only metastable. The
difference of free energy of the typical localized solution and the uniform
state is the configurational entropy times the absolute temperature, as
confirmed using replica methods\cite{mezard.1999}.

To estimate the interaction energies, Xia and Wolynes\cite{xia.2000} pointed
out that their total, at the Kauzmann temperature $T_{K}$ where the
configurational entropy vanishes, must equal the localization cost $%
T_{K}S_{loc}$. Therefore, if a typical molecule has z nearest neighbors a
local interaction must contribute a term $v_{int}=(1/z)T_{K}S_{loc}$ on the
average. The localization entropy cost, in the free energy functional,
depends logarithmically on the amount of space each molecule can move in
while being encaged: $S_{loc}=\frac{3}{2}k_{B}\log (\alpha _{L}/\pi e)$
where $\alpha _{L}$ is the inverse square of the so-called Lindemann ratio
of the r.m.s. vibrational amplitude in the glass to the intermolecular
spacing. The Lindemann ratio is predicted by detailed microscopic
calculations\cite{singh.1985,stoessel.1984,hall.2003} and can be measured by
neutron scattering where it determines the height of the long-time plateau.
The Lindemann ratio only weakly depends on the intermolecular potential and
is of order $\frac{1}{10}$ near $T_{g}$. Thus we see, the microscopic RFOT
theory suggests $v_{int}$ should be nearly the same in units of $k_{B}T_{K}$
for all molecular glass formers made of spherical particles and predicts its
value. This near universality of the interaction per molecular unit allows
the RFOT theory to make quantitative predictions in a wide variety of
substances of many measured quantities characterizing glassy motion such as
the typical barriers near $T_{g}$\cite{xia.2000}, the degree of
nonexponentiality\cite{xia.2001} and the correlation length.

Once the density functional approach has generated aperiodic minima, the
dynamics of escape from a given one resembles very much the dynamics of
overturning regions of a random field Ising magnet (RFIM) in a field. The
free energy difference on a site predicted by the density functional acts
like a magnetic field of magnitude $TS_c(T)$ in the RFIM. This quantity
fluctuates, and the magnitude of its fluctuations are of the order $\sqrt{%
k_BT\Delta C_p}$ where $\Delta C_p$ is the configurational heat capacity of
the fluctuating region. The interaction between a pair of sites in the RFIM
analogy is $v_{int}$, which is already computed. Using this quantitative
analogy, RFOT theory computes the typical escape barrier and the
fluctuations of the barriers near $T_K$.

To estimate the barriers for escape, we can characterize the shape of a
reconfiguring region by the number of contiguous sites N that are rearranged
and the number of surface interactions that are broken, b. Near to $T_K$,
the regions that dynamically reconfigure should be compact because this
involves losing the smallest number of favorable interactions, b, while
gaining the same configurational entropy proportional to N.

Maximal compactness implies a roughly spherical shape giving a free energy
cost 
\begin{equation}
\Delta F(N)=-TS_{c}N+v_{int}\frac{z}{2}4\pi \left( \frac{N}{4\pi /3}\right)
^{2/3}  \label{eq.1}
\end{equation}%
yielding a barrier that diverges in three dimensions like $S_{c}^{-2}$. This
result is modified by the multiplicity of aperiodic states. In analogy to
Villain's treatment of the RFIM\cite{villain.1985}, near $T_{K}$ the
interface of the reconfigured region between any two aperiodic patterns will
be wetted by other specific aperiodic minima that better match the two
abutting regions than they do already. This lowers the surface energy term
to now scale like $N^{1/2}$. This form for the mismatch energy restores the
scaling relations near $T_{K}$\cite{kirkpatrick.1989}. This wetting effect
has been shown to be connected with additional replica symmetry breaking in
the interface in replica instanton calculations\cite{dzero.2005,franz.2005b}%
. Wetting cannot occur at short ranges so the scale of this mismatch term
still follows from $v_{int}$. In this way the observed Vogel-Fulcher scaling
near $T_{K}$ is predicted, $\Delta F^{\ddagger }\propto S_{c}^{-1}$. The
numerical proportionality coefficient can be computed from the microscopic
value of $v_{int}=\frac{1}{z}\frac{3}{2}k_{B}T_{K}\log \frac{\alpha _{L}}{%
\pi e}$, and the result is a universal multiple of $k_{B}T_{K}$. This
resulting prediction of absolute activation barriers agrees well with
experimental results for 44 substances\cite{stevenson.2005,lubchenko.2003},
a typical deviation being less than 20\%.

The compact shape of the CRR and the Vogel-Fulcher behavior are
asymptotically correct near $T_K$. We now emphasize that away from $T_K$ the
CRR need not be compact. Other shapes have an entropy advantage: Although
the sphere (for which $b=\frac{z}{2}4\pi \left( \frac{N}{4\pi /3}
\right)^{2/3}$) is unique, there are many contiguous structures with other
shapes. Increased temperature favors these more ramified shapes as CRR's.
Contiguous shapes are called lattice animals\cite{stauffer.1978}. Much work
has gone into enumerating lattice animals because of their importance in
problems such as in percolation\cite{stauffer.1985}, Yang-Lee zeros\cite%
{yang.1952}, etc. Klein and Unger\cite{unger.1984} emphasized that near a
spinodal of an ordinary first order transition the dominant nuclei should be
lattice animals characteristic of clusters at the percolation threshold. We
now extend this reasoning to random first order transitions.

Accounting for the multiple possible shapes of a CRR we can write the free
energy of moving any cluster of N sites with b boundary interactions as 
\begin{equation}
\Delta F(N,b)=-TS_{c}N+v_{int}b-k_{B}T\log (\Omega (N,b))  \label{eq.2}
\end{equation}%
where $\Omega (N,b)$ is the number of lattice animals of given N and b. For
a given N the most numerous shapes are percolation-like. When these states
dominate we can use enumeration studies near the percolation limit to
evaluate $\Omega (N,b)$. In percolation clusters, Leath\cite{leath.1976}
determined that for large N, 
\begin{equation}
\Omega _{perc}(N,t)\sim \left( \frac{\left( \alpha +1\right) ^{\alpha +1}}{%
\alpha ^{\alpha }}\right) ^{N}\exp \left( -\frac{N^{2\phi }}{2B^{2}}\left(
\alpha -\alpha _{e}\right) ^{2}\right)  \label{eq.leath}
\end{equation}%
Here, $\alpha =t/N$, and $t$ is the number of unoccupied sites bounding the
occupied cluster. We will take $\phi $ to have its mean field value of 1/2.
B is a lattice dependent constant. $B=1.124$ (for the face centered cubic
lattice) follows from fitting the Leath formula to numerics calculated by
Sykes et al.\cite{sykes.1976} for $N\leq 9$. The mean value of $t/N$
approaches $\alpha _{e}=(1-p_{c})/p_{c}$ for large N at the percolation
threshold, $p_{c}$ ($p_{c}=0.198$\cite{sykes.1976} for the FCC lattice).

Ideally we would like to evaluate the needed percolation quantities for
random close packed lattices. There is an ambiguity here as to the
definition of ``contact''. Clearly spheres need not precisely touch (as in,
say, percolation conductivity experiments), but rather their surfaces may be
separated by at most a Lindemann length in order to be called connected.
While the parameters for this continuum percolation problem are not
available, they can be easily estimated since the percolation quantities
primarily depend on the near neighbor connectivity. The number of neighbors
in the rcp lattice is roughly the same as the face centered cubic (FCC);
thus it is reasonable to use parameters for an FCC close packed lattice of
spheres.

The number of bonds, b, is related to Leath's t. For the simple cubic
lattice, $\frac{<b>}{<t>}=1.67$\cite{cao.1992}. This ratio is linear in
coordination number, z, so we can again find the value for the rcp lattice.
Thus, following from equation \ref{eq.2}, 
\begin{equation}
\Delta F(N,t)=-TS_{c}N+v_{int}1.68\frac{z}{z_{SC}}t-k_{B}T\log (\Omega
_{perc}(N,t))  \label{eq.4}
\end{equation}%
To find the most probably escape route we need to find the most probable t
as a function of N. If we minimize equation \ref{eq.4} with respect to $t$
using equation \ref{eq.leath} for $\Omega $, we find that the most probable
value of $t$ is $\bar{t}=\bar{\alpha}N$ where $\bar{\alpha}=3.10$. With this
most probable value, $\Omega _{perc}$ becomes simply $\Omega _{perc}=\lambda
^{N}$ where $\lambda =7.64$. Thus each term in equation \ref{eq.4} is now
proportional to N, and the free energy becomes. 
\begin{equation}
\Delta F(N)=\left( -TS_{c}+v_{int}1.68\frac{z_{FCC}}{z_{SC}}\bar{\alpha}%
-k_{B}T\log (\lambda )\right) N  \label{eq.5}
\end{equation}

Apart from $S_c$ each term in this expression follows from a microscopic
calculation. The nature of the profile therefore depends on the
configurational entropy. Clearly the free energy for percolative CRR's
either monotonically increases or decreases with N. If the free energy
profile increases with N, a reconfiguration event due to a percolation
cluster is impossible, so a more compact structure that will eventually
become stable for large N provides the dominant reconfiguration route. If F
decreases with N for the percolation shape, no barrier at all should be
observed. Clearly the change of behavior of $\Delta F(N)$, from increasing
to decreasing with N, signals a crossover to non-activated dynamics. Taking $%
v_{int}$ to be entirely entropic and putting the determined RFOT values of $%
v_{int}$ into equation \ref{eq.5} yields. 
\begin{equation}
\begin{split}
\Delta F(N)&=(-TS_c+k_BT(3.20-1.91))N \\
&=-T(S_c-k_B1.28)N \\
\end{split}%
\end{equation}
We see barrier-less reconfiguration events occur at a critical
configurational entropy, $S_c^{perc} = 1.28k_B$ if we neglect the mean field
softening effects. Using the thermodynamic relation, $S_c(T)=S_\infty
(1-T_K/T)$ where $S_\infty $ is given by $\Delta C_p(T_g)T_g/T_K$, RFOT
theory thus predicts the crossover transition temperature, $T_c^{perc}$. 
\begin{equation}
\frac{T_c^{perc}}{T_K}=\left(1-\frac{S_c^{perc}}{\Delta C_p}\frac{T_K}{T_g}%
\right)^{-1}  \label{eq_Tperc}
\end{equation}
The bigger $\Delta C_p$ is, the closer $T_c^{perc}$ will be to $T_K$; more
``fragile'' liquids with larger $\Delta C_p$ have a smaller activated range,
while a broader range for activated transport applies for stronger liquids
with smaller $\Delta C_p$. A similar trend is predicted for the mean field
crossover based on detailed microscopic calculations for fluids with network
network structure by Hall and Wolynes\cite{hall.2003} who suggest an entropy
at the higher mean field crossover of $S_c(T_A)=2.0k_B$. Including the
softening of $v_{int}$ expected as this mean field transition is approached
lowers the estimate of the percolation point. The amount of lowering is
uncertain, however, because simultaneous with the softening a broadening of
the interface is expected, thus effectively reducing the possible entropy
gain from shape fluctuations. RFOT theory indicates that at the same
configurational entropy level, on the average, the nature of transitions
will be the same in fragile and strong liquids. As in the RFOT theory of the
non-exponentiality parameter $\beta$\cite{xia.2001}, fluctuations in the
driving force depend on $\Delta C_p$ explicitly and should be included in
equation \ref{eq.1}. Thus fast and slow CRR's would have somewhat different
shapes (faster being more ramified generally since their entropy is higher).
We do not discuss this effect further here, however.

\begin{figure}[tbp]
\includegraphics[width=0.45\textwidth]{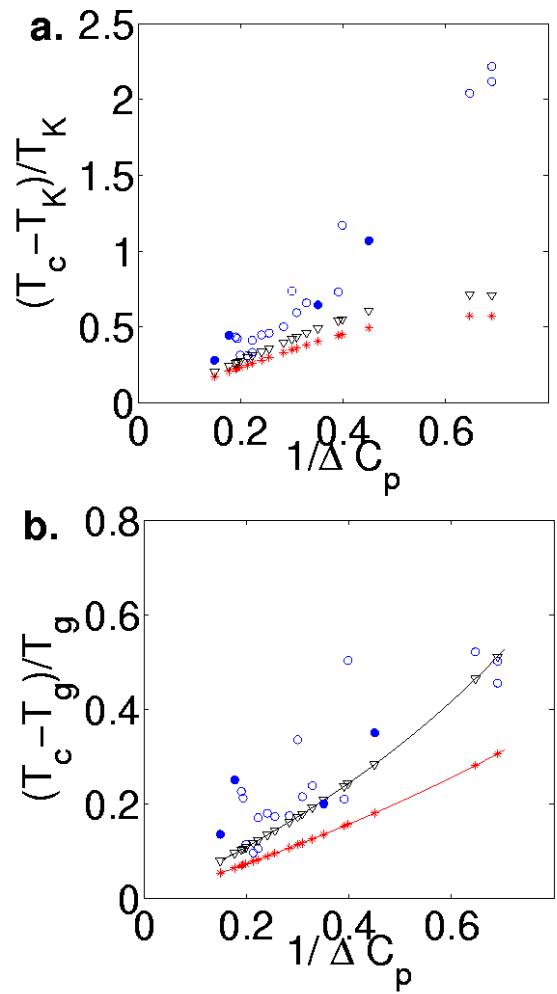}
\caption{\textbf{a}. Predictions for $(T_c^{string}-T_K)/T_K$ (depicted as
*stars) and $(T_c^{perc}-T_K)/T_K$ (depicted as triangles). The
experimentally derived crossover temperatures, $(T_c^{exp}-T_K)/T_K$, from
those materials collected by Novikov and Sokolov\protect\cite{novikov.2003},
are shown as circles with the dark circles referring to polymers. In all
cases the values for the Kauzmann temperature, $T_K$, were taken from the
correlation $T_K=T_g(1-16/m)$ found in a paper by B\"{o}hmer and Angell%
\protect\cite{bohmer.1992}. \textbf{b}. Same as for \textbf{a}. except a
plot of $(T_c-T_g)/T_g$ instead of $(T_c-T_K)/T_K$. The conversion ratio $%
T_K/T_g$ was set through $S_c(T_g)=\Delta C_p(T_g-T_K)/T_K=0.79k_B$. For
both plots the $\Delta C_p$ values for the materials were determined from
their m values through the correlation $m=20.7\Delta C_p$ discussed in
Stevenson and Wolynes\protect\cite{stevenson.2005}, where $\Delta C_p$ is,
from RFOT theory, the heat capacity discontinuity at $T_g$ per independently
moving unit, or ``bead.''}
\label{fig.2}
\end{figure}

We see that the counting problems for percolation clusters are not all that
different from those relevant for strings. We can re-do the crossover
transition argument for purely string-like objects. The number of broken
interactions of a string scales with length, $N(z-2)$, as does the shape
entropy of a string, $\log (\Omega )=N\log (z-5)$. $(z-5)$ represents the
number of directions a string can take that excludes backtracking on top of,
or directly next to, the previous particle. The next neighbor position is
disallowed because, if occupied, the cluster becomes compact. Putting these
coefficients into the free energy of growing a string of length N, we find
that string growth will become the down hill route of escaping a minimum at
an entropy of $S_{c}^{string}=1.13k_{B}$. This results in a crossover
temperature 
\begin{equation}
\frac{T_{c}^{string}}{T_{K}}=\left( 1-\frac{S_{c}^{string}}{\Delta C_{p}}%
\frac{T_{K}}{T_{g}}\right) ^{-1}
\end{equation}%
a bit lower than predicted by percolation. In Figure \ref{fig.2}a we plot
the predicted $T_{c}^{string}$ and $T_{c}^{perc}$ versus $1/\Delta C_{p}$
for various liquids. Crossover temperatures from activated to non-activated
dynamics have been determined by Stickel plot analysis\cite{stickel.1996}.
Experimental crossover temperatures for 21 substances obtained in this way
by Novikov and Sokolov\cite{novikov.2003} are also plotted in the figure.
Some of the outliers are polymers for which other slowing effects compound
simple RFOT results(). Uncertainty in the determination of $T_{K}$ for very
strong liquids probably is a source of the discrepancy between the theory
and experiment for these latter substances. We have also plotted $%
(T_{c}-T_{g})/T_{g}$ in figure \ref{fig.2}b. According to RFOT theory the
entropy at $T_{g}$ is $S_{c}(T_{g})=\Delta C_{p}(T_{g}-T_{K})/T_{K}=0.79k_{B}
$. The value of $S_{c}(T_{g})=0.79k_{B}$ gives a glass transition occurring
at $10^{10}P$, as was used in the experimental papers (see figure \ref{fig.4}%
). The quantitative agreement of the experimental crossover temperatures and
the present predictions of the string and percolation transitions is
striking.

To predict quantitatively the barriers and the typical shapes of
reconfiguring regions at temperatures between $T_c$ and $T_K$ we must find a
suitable analytic form of $\Omega (N,b)$ for all relevant values of N and b.
Surface roughening theories give predictions of $\Omega (N,b)$ valid only
for nearly spherical objects\cite{chui.1976}, and would be useful only near $%
T_K$. The percolation theory gives an explicit form of $\Omega$ valid only
for the most populous ramified, fractal shapes that dominate near the
crossover.

\begin{figure}[tbp]
\includegraphics[width=0.48\textwidth]{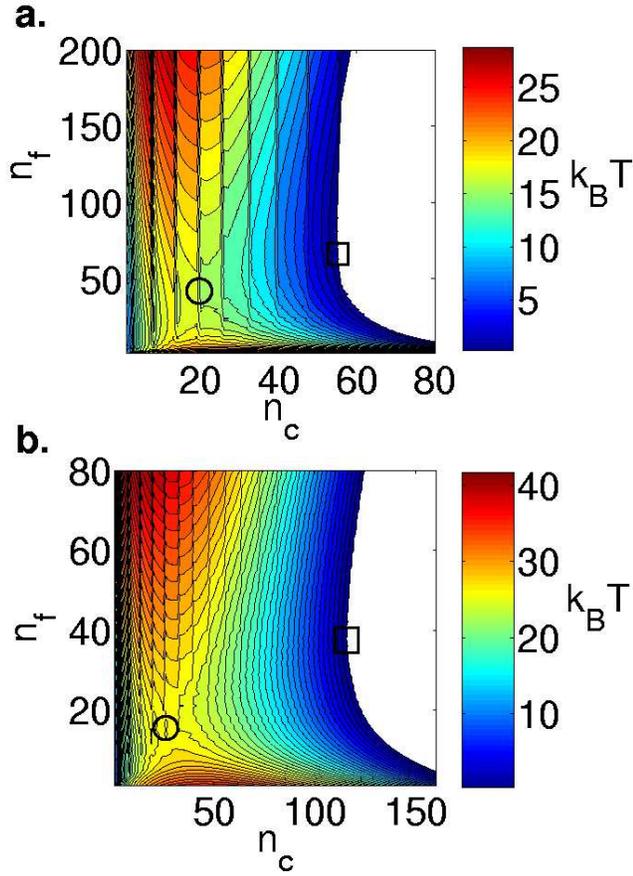}
\caption{Two dimensional free energy profiles as functions of the number of
sites in the core, $n_c$, and the number in the fuzzy halo, $n_f$, \textbf{a}%
. near $T_c^{string}$ and \textbf{b}. near $T_g$. The sidebar is in units of 
$k_BT$ with the contours lines corresponding to intervals of $1k_BT$. The
circles indicate the location of the typical transition state. The squares
indicate a fully reconfigured region.}
\label{fig.3}
\end{figure}

\begin{figure}[tbp]
\includegraphics[width=0.48\textwidth]{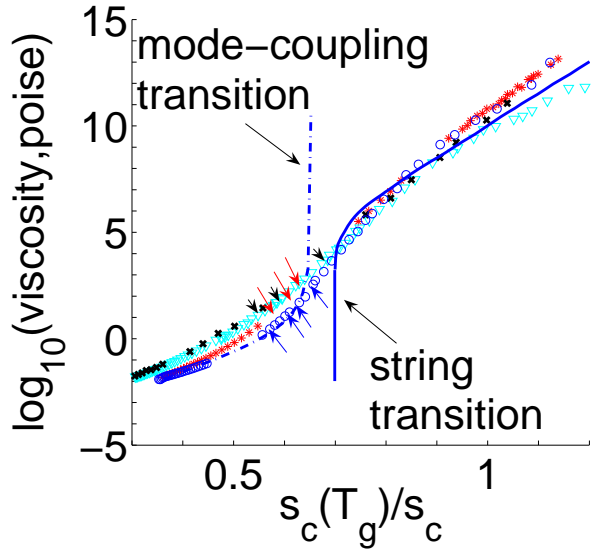}
\caption{A comparison of experimental viscosity barriers with the barriers
predicted from the fuzzy sphere model (solid line). Salol data\protect\cite%
{stickel.1996} are represented with circles, Propylene Carbonate data%
\protect\cite{stickel.1996} with crosses, O-Terphenyl data\protect\cite%
{laughlin.1972,cukierman.1973} with stars, and alpha-Phenyl-O-Cresol data%
\protect\cite{laughlin.1972,cukierman.1973} with triangles. An experimental
mode coupling fit to salol (Hinze et al.\protect\cite{hinze.2000}) is shown
with a dot-dashed line. Experimentally derived values of the entropy at the
crossover transitions\protect\cite{novikov.2003} are shown with arrows. The
free energy barriers were placed on the $log_{10}(viscosity)$ curve by
setting $\Delta F^{\ddagger} =0$ to correspond with the large T experimental
value of 1 centipoise for the viscosity. At the measured calorimetric glass
transition, corresponding to a cooling rate of $10\,^{\circ}/min$, the
viscosity of the materials is about $10^{10}P$. This value was used to
determine the theoretical value $S_c(T_g)=0.79k_B$.}
\label{fig.4}
\end{figure}

\begin{figure}[tbp]
\includegraphics[width=0.4\textwidth]{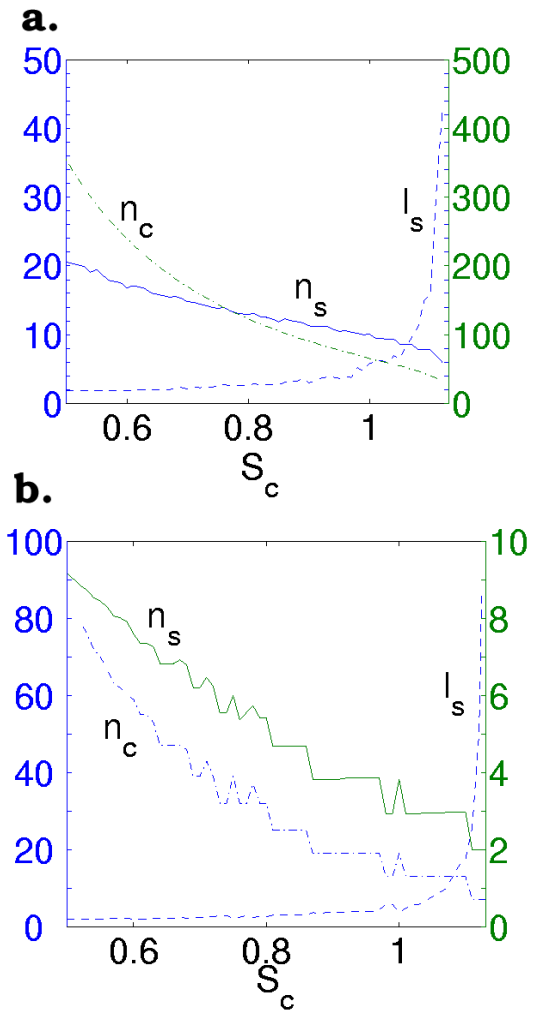}
\caption{Predictions for the fuzzy sphere shape characteristics are shown as
functions of the configurational entropy for the final state and the
transition state. $n_c$ (dashed-dotted line) is the number of particles in
the core, $n_s$ (solid line) is the number of strings and $l_s$ (dashed
line) is the typical length of a string. \textbf{a}. The final state: $n_c$
(dashed-dotted line) uses the axes on the right while $n_s$ (solid line) and 
$l_s$ (dashed line) use the axes on the left. \textbf{b}. The transition
state: here $n_s$ (solid line) uses the axes on the right while $n_c$
(dashed-dotted line) and $l_s$ (dashed line) use the axes on the left. The
sizes and lengths are given in terms of the number of particles. The strings
in the halo are nearly random flight chains giving the physical dimensions
in figure \protect\ref{fig.6}.}
\label{fig.5}
\end{figure}

\begin{figure}[tbp]
\includegraphics[width=0.48\textwidth]{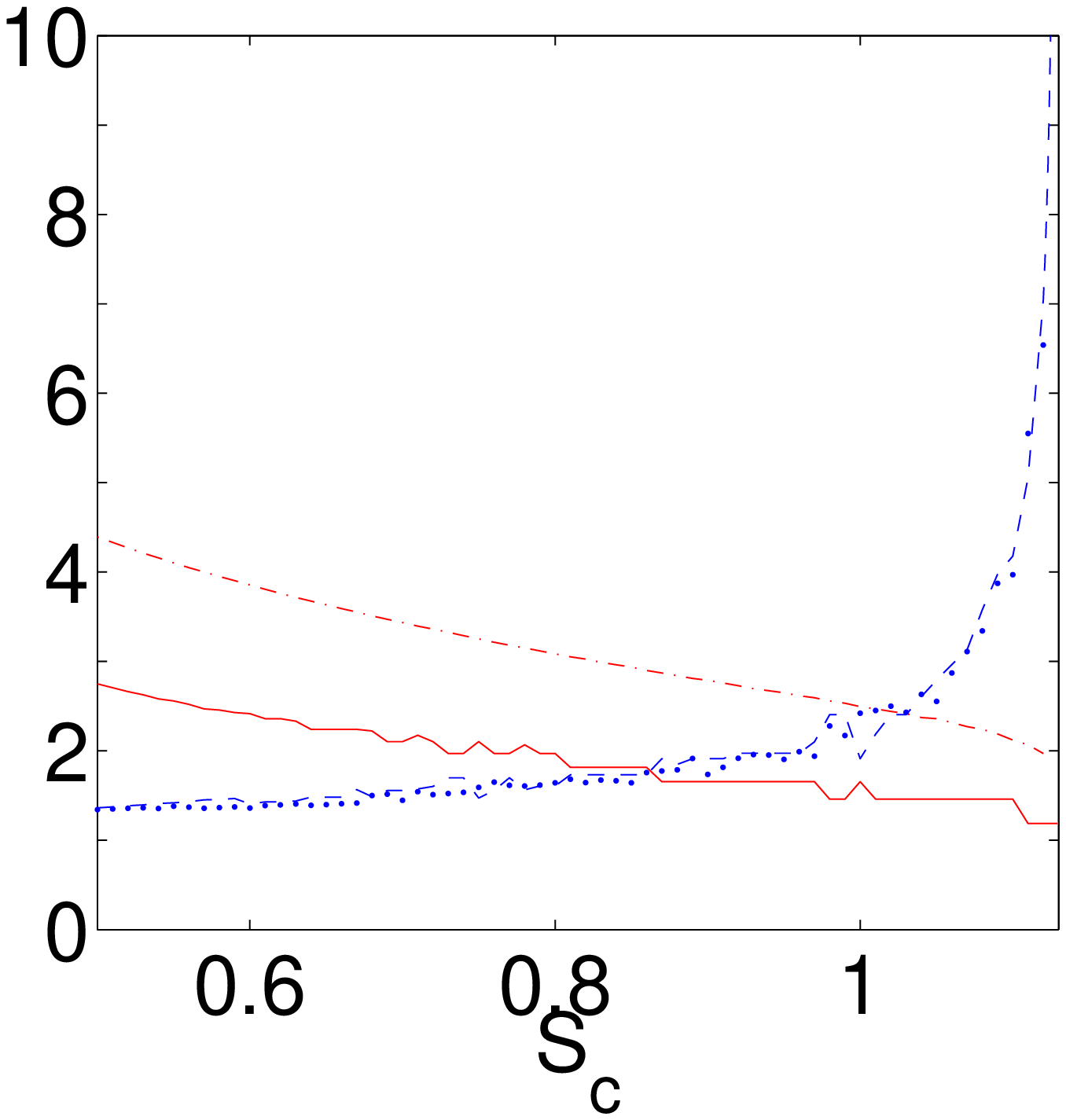}
\caption{The radius of the core, $R_c$, at the transition state (solid line)
and at the final state (dashed-dotted line). Also, the radius of the stringy
halo, $R_s$, at the transition state (dashed line) and the final state
(dotted line). The radii are given in terms of the number of particles.}
\label{fig.6}
\end{figure}

We use a reasonably effective, but unabashedly approximate, treatment of the
animal counting problem to interpolate between these limits. We describe the
reconfiguring region as a ``fuzzy sphere,'' an object with a spherical core
of $n_c$ particles, but surrounded by a ramified, but connected, halo of $%
n_f $ particles. If we let the core size, $n_c$, go to zero we are left with
only an extended object. Conversely, if we let the halo vanish then we have
only a sphere. We would like to describe the halo as a percolation cluster,
but all the relevant geometric quantities are not precisely known. We
therefore model the halo as a set of strings of particles extending from the
surface of the central core. In this way we can explicitly determine the
entropic contribution of a halo of $n_s$ strings.

For $S_c$ between the extremes of $0$ (at $T_K$) and $S_c^{string}$ we can
now compute the full activation free energy profile for CRR's pictured as
fuzzy spheres. 
\begin{widetext}
\begin{equation} 
\begin{split} 
\Delta F(n_c,n_f,n_s)=&v_{int}\frac{z}{2}\left(\frac{4\pi /3}{n_c}\right)^{1/6}\left(4\pi \left(\frac{n_c}{4\pi /3}\right)^{2/3}-n_s 
\right)+v_{int}(z-2)n_f \\ 
&-TS_c(n_c+n_f)-k_BTlog(\Omega (n_c,n_f,n_s)) 
\end{split} 
\end{equation} 
\begin{equation} 
\Delta F(n_c,n_f)=-log\left(\sum_{n_s}exp(-\Delta F(n_c,n_f,n_s))\right) 
\end{equation} 
\end{widetext}
The full expression for the shape entropy of a fuzzy sphere, $\Omega$, is
found in the supplementary material. This estimate accounts for the excluded
volume between the strings via Flory theory\cite{flory.1953}. Figures \ref%
{fig.3}a and \ref{fig.3}b show contour plots of the free energy at a
configurational entropy value near the dynamic crossover and near the glass
transition respectively. The saddle points on these free energy surfaces
describe transition state ensembles for activated reconfiguration events.
The resulting barrier depends universally on the configurational entropy and
is plotted as a function of $S_c(T_g)/S_c(T)$ in figure \ref{fig.4}.
Superimposed on the graph are the experimental viscosity barrier for several
glass forming liquids of varying fragility and of known entropy and heat
capacity. The universality is clearly confirmed (In these plots the
calorimetrically determined $T_g$'s were used for calibration, not the
viscometric values!). The barrier clearly depends linearly on $1/S_c$ for $%
S_c<S_c^{string}$ consistent with the asymptotic RFOT analysis, but as the
critical value of the configurational entropy, $S_c^{string}$, is approached
the activation barrier rapidly decreases, dropping to zero at $S_c^{string}$%
. We have also included in the plot the experimental mode coupling fit to
the viscosity for salol\cite{hinze.2000}. The symmetry is striking; the mode
coupling theory fits the dynamic transition from above, while the current
argument predicts its emergence from below.

As well as giving the barriers, this calculation suggests the dominant
routes the liquid takes to reconfigure and the final form of the CRR's in
different entropy regimes. The final states are broadly distributed in shape
as shown via the broad $1k_{B}T$ contour in the plots. We define the
characteristic final shape as the one with the smallest core. We show
examples of the final shape expected near $T_{g}$ and near the crossover
temperature in figure \ref{fig.1}. Figures \ref{fig.5} and \ref{fig.6} show
how the characteristic scales of the transition states and final
reconfigured regions change with configurational entropy. Near $T_{g}$ the
shapes are mostly spherical with just a small fraction of the particles in
the stringy halo and have sizes consistent with the previous XW estimate. A
typical protuberance on the compact core near $T_{g}$ is only 2 particles
long. Near $T_{c}^{string}$, however, the core size becomes very small while
the strings lengthen dramatically. This growth occurs for both the
transition state and the final state.

The string lengths near $T_{c}^{string}$ are larger than what is usually
reported in simulation or in microscopy studies. This apparent discrepancy
arises from a kinetic effect which we describe as follows: Though the free
energy barrier for creating a string approaches zero at $T_{c}^{string}$,
the actual time to construct a string not only remains finite but indeed
grows with the length of the string. The barrier to create a new string is
somewhat larger than to extend an old one. Because of this, the growth/death
of a string, generally takes place particle by particle on the microscopic
time scale, and should be a diffusive process, with growth time $\tau
_{s}=\tau _{micro}^{0}l_{s}^{2}$. Here, $\tau _{micro}^{0}$ is a typical
vibrational time scale, i.e.the time for a particle to explore its cage.
When $\tau _{s}$ becomes comparable to the time for another activated event
to occur in the immediate vicinity of the string, $\tau _{\alpha }/l_{s}$,
the growth of the original string will be perceived to have been
interrupted. Here, $\tau _{\alpha }=\tau _{micro}^{0}e^{F^{\ddagger }/k_{B}T}
$. This finite growth time gives a maximum limit for the length of strings:%
\begin{equation}
l_{s,max}^{3}=e^{F^{\ddagger }/k_{B}T}.
\end{equation}%
Larger strings will be interrupted, or \textquotedblleft
incoherent,\textquotedblright\ since an activated event occurs along the
string. Using the fuzzy sphere model, the minimum barrier corresponds to a
core region of size $7$. This gives an $F^{\ddagger }$ consistent with what
Novikov and Sokolov\cite{novikov.2003} call the \textquotedblleft
magic\textquotedblright\ relaxation time for the crossover and a length $%
l_{s.max}\cong e^{14/3}\cong 108$. This is larger than the lengths usually
quoted from simulations currently made on long time scales, but the rapid
variation of $F^{\ddagger }$ and $l_{s}$ near the string transition makes
this result rather sensitive to modeling details. Important is that there is
a natural cut off length of kinetic origin that causes $T_{c}^{string}$ to
be a crossover scale and not a sharp transition.

We see that the random first order transition theory predicts the CRR's in
glassy liquids are compact, nearly spherical objects in the deep supercooled
region, but that in the moderately supercooled region, near the mode
coupling transition, the CRR's are predicted to become non-compact, extended
string-like objects. The crossover temperature is entropically controlled.
This prediction of RFOT theory is confirmed by experiment.

\bigskip

\begin{acknowledgments}
Work at UCSD was supported by NSF grant CHE0317017. J. Schmalian was
supported by the Ames Laboratory, operated for the U.S. Department of Energy
by Iowa State University under Contract No. W-7405-Eng-82 (J.S.).
\end{acknowledgments}

\bigskip

\appendix
\noindent\textbf{Supplementary Material:} The number of ways a fuzzy sphere
with $n_c$ particles in the central core, $n_f$ particles in the stringy
halo, and $n_s$ strings can be arranged is $\Omega(n_c,n_f,n_s)$. The
entropy, $log(\Omega)$, of a fuzzy sphere can be broken up into three
components 
\begin{equation}
\Omega(n_c,n_f,n_s)=e^{S_{pos}+S_{dis}+S_{grow}}
\end{equation}
$S_{dis}$ is the number of ways one can distribute $n_f$ particles in $n_s$
strings ensuring each string has at least one particle. 
\begin{equation}
e^{S_{dis}}=\frac{(n_f-1)!}{(n_f-n_s)!(n_s-1)!}
\end{equation}
The positional entropy, $S_{pos}$, accounts for the number of ways you can
position $n_s$ strings on a surface of area $n_{surf}=4\pi \left(\frac{n_c}{%
4\pi /3}\right)^{2/3}$. It is necessary to account for the excluded volume
of the strings by noting that a string not only excludes its own position,
but also $\tilde{z}=z/2$ of its nearest neighbors as well. Each string
placed on the surface further reduces the number of available positions by $%
\tilde{z}+1$. Using this we can find a form of $S_{pos}$ by adding particles
one at a time. 
\begin{widetext}
\begin{equation}
e^{S_{pos}}=\frac{1}{n_s!}(n_{surf})(n_{surf}-(\tilde{z}+1))(n_{surf}-2(\tilde{z}+1))\cdots(n_{surf}-(n_s-1)(\tilde{z}+1))
\end{equation}
\end{widetext}
The growth entropy $S_{grow}$ accounts for the number of configurations the
strings can take on as they lengthen. We calculate this excluded volume
effect on the entropy of the strings using Flory's method\cite{flory.1953}.
We imagine the fuzzy halo as made up of $n_f/n_s$ concentric shells each
containing one particle for each of the $n_S$ strings. If we place particles
one at a time, filling up each shell before starting a new one, then $f_{ij}$
is the expectation value that a given cell is occupied or part of the
excluded volume of the $i-1$ previously placed particles in the $j^{th}$
shell. $f_{ij}$ can be estimated by the fraction of the current shell taken
up by the excluded volume of the other particles. 
\begin{equation}
f_{ij}=\frac{\tilde{z}(j-1)}{4\pi \left(\left(\frac{n_c}{4\pi /3}%
\right)^{1/3}+i-1\right)^2}
\end{equation}
Thus, including the $n_s$ particles per shell in all $n_f/n_s$ shells we get 
\begin{equation}
e^{s_{grow}}=(z-5)^{n_f-n_s}\prod_{i=2}^{n_f/n_s}\prod_{j=1}^{n_s}(1-f_{ij})
\end{equation}

\newpage ~

\bibliography{/home/jake/latex_files/jakes_biblio.bib}

\end{document}